# Friedel Oscillation-Induced Energy Gap Manifested as Transport Asymmetry at Monolayer-Bilayer Graphene Boundaries

Kendal W. Clark[1], X.-G. Zhang[1], Gong Gu[2], Jewook Park[1], Guowei He[3], R. M. Feenstra[3], and An-Ping Li[1*]

[1] Center for Nanophase Materials Sciences, Oak Ridge National Laboratory, Oak Ridge, Tennessee 37831, USA

[2] Department of Electrical Engineering & Computer Science, University of Tennessee, Knoxville, TN 37996, USA

[3] Department of Physics, Carnegie Mellon University, Pittsburgh, PA 15213, USA

* Email:apli@ornl.gov

We show that Friedel charge oscillation near an interface opens a gap at the Fermi energy for electrons with wave vectors perpendicular to the interface. If the Friedel gaps on two sides of the interface are different, a nonequilbrium effect – shifting of these gaps under bias – leads to asymmetric transport upon reversing the bias polarity. The predicted transport asymmetry is revealed by scanning tunneling potentiometry at monolayer-bilayer interfaces in epitaxial graphene on SiC (0001). This intriguing interfacial transport behavior opens a new avenue towards novel quantum functions such as quantum switching.

## I. INTRODUCTION

Friedel charge density oscillations near defects such as impurities and boundaries are ubiquitous in metallic materials [1]. They usually have little effect on electron transport in metals, however, because the large electron density in a metal dwarfs the Friedel oscillation, rendering the perturbation on the electronic structure caused by such an oscillation negligible. The situation can be different in materials such as graphene and topological insulators, where the electron density is often low and the Coulomb interaction can therefore be important [2, 3]. Friedel oscillations in graphene [4-6] and topological insulators [7-9] have been studied both theoretically and experimentally in equilibrium states, where, similar to metals, the Fermi momentum $k_F$ is well-defined and such oscillations are characterized by the wave vector $Q = 2k_F$. However, in transport experiments on materials with low electron densities, the system can be easily driven far from equilibrium and the Friedel oscillations become voltage dependent [10]. Low electron density and strong electron-electron interaction in such systems, combined with the sensitivity of the Friedel oscillation to nonequilibrium effects, should lead to strong influence on transport properties. Little effort has been spent on exploring the effects of Friedel oscillations on electron transport.

Here we show that Friedel oscillation can profoundly impact electron transport across an interface: The electrostatic potential due to the Friedel oscillation couples the right- and left-going waves near the Fermi energy and opens an energy gap for normally incident electrons, representing an extra energy cost for electron transmission across the interface. Because of the dependence of the Friedel oscillation period on the bias voltage, these gaps can manifest themselves as asymmetric electrical transport across the interface if the gaps on both sides are different. Such a transport asymmetry is experimentally demonstrated in our scanning tunneling potentiometry measurements across a monolayer-bilayer boundary in epitaxial graphene formed on SiC (0001).

We will first develop a general theory of the Friedel energy gap and the transport asymmetry across a boundary due to such a gap in Section II. In Section III we present the experimental procedure for measuring the transport asymmetry in graphene. The measured results are compared to theory in Section IV and are discussed in Section V.

## II. THEORY

### A. Friedel energy gap

We first consider the effect of a charge density oscillation of constant amplitude on an electron at the Fermi energy in a system that has a low electron density where each lattice site is occupied by $n = n_\uparrow + n_\downarrow << 1$ electrons, with $\uparrow$ and $\downarrow$ indicating spin. The onsite Coulomb energy for an itinerant electron, which contributes an additional term $\delta n << n$ to the charge on the lattice site, is $U(n_\uparrow + \delta n_\uparrow)(n_\downarrow + \delta n_\downarrow) - Un_\uparrow n_\downarrow \approx \frac{1}{2}Un\delta n$, where $U$ is the Hubbard energy. Within the mean-field approximation $\langle n \rangle = \Omega\rho(x)$ for a charge density $\rho(x) = \rho_0 + \rho_1\cos(Qx + \phi)$, where $\Omega$ is the unit cell volume (or area for 2D systems), $\rho_0$ the average charge density, $\rho_1$ the amplitude of the charge density oscillation, and $Q$ the wave vector of the charge density oscillation. The Coulomb energy due to the Friedel oscillation means that there is an extra Coulomb term in the Hamiltonian within the mean-field approximation,

$$H_1 = \frac{1}{2}U\Omega\rho_1\sum_i c_i^+ c_i \cos(Qx_i + \phi), \qquad (1)$$

where $c_i^+$ is the creation operator at lattice position $x_i$, and $\varphi$ is the phase of the oscillation. One can verify that Eq. (1) reproduces the Coulomb energy of an itinerant electron if $\langle c_i^+ c_i \rangle = \delta n$. Consider a pair of states, one right-going along the direction of the charge density oscillation and the other left-going. Suppose the wave functions of the two states without $H_1$ are $|k_x = Q/2 + q\rangle = e^{i(Q/2+q)x}/\sqrt{\Omega}$ and $|k_x = -Q/2 + q\rangle = e^{-i(Q/2-q)x}/\sqrt{\Omega}$, respectively, and their energies are $E(Q/2 \pm q) = E(Q/2) \pm cq$, which, with $c$ being the group velocity, is a good approximation for small $q$ if $E(Q/2)$ is sufficiently away from a band edge. $H_1$ couples these two states. By first-order perturbation, the eigenstates are standing waves with energies

$$E\left(\frac{Q}{2} \pm q\right) = E\left(\frac{Q}{2}\right) \pm \frac{1}{4}\sqrt{(U\Omega\rho_1)^2 + (4cq)^2} \qquad (2)$$

Therefore, a gap is opened at energy $E(Q/2)$, which we refer to as the Friedel gap ($G_F$) hereafter. At equilibrium, $Q = 2k_F$ and the gap is located at the Fermi energy, as schematically sketched in Fig. 1a. The Friedel gap is different than the charge density wave condensate in low-dimensional systems. The latter is usually the result of electron-phonon coupling, where a gap at the Fermi energy is formed from lattice distortion due to, e.g., the Peierls instability [11]. In the case of Friedel oscillation, the gap is formed simply due to the electron-electron interaction.

## B. Transport asymmetry

In a nonequilibrium state, i.e. under a bias $V$, transport asymmetry arises from an intuitive result that the Friedel gap on the transmitted side moves outside the transport window. This effect is intuitive because the transmitted electron wave functions do not form interference patterns that contribute to the Friedel oscillation. Indeed, a more rigorous consideration using nonequilibrium Green's function showed that the period of the Friedel oscillation increases from its equilibrium value [10]. For bands with $c > 0$, this in turn shifts the Friedel gap downwards relative to the equilibrium Fermi energy. Numerical modeling found the bias voltage dependence of the Friedel period to satisfy the condition [10],

$$E\left(\frac{Q}{2}\right) = E(k_F) - \frac{1}{2}|eV| \qquad (3)$$

which with a linear dispersion leads to $Q = 2k_F - |eV/c|$. The center of the Friedel gap therefore moves from the chemical potential [12] $\mu$ to $\mu - |eV/2|$. There exists a saturation voltage, $V_c = G_F/e$, beyond which the Friedel gap moves so much that it is completely below the equilibrium chemical potential $\mu = E(k_F)$.

At an interface under a bias voltage $V$, the chemical potentials of the two sides, $\mu_L$ and $\mu_R$, differ by $eV$. Naively, if one neglects the voltage dependence of the Friedel oscillation period, the Friedel gaps would move rigidly with the chemical potentials (as schematically shown in the upper panel of Fig. 1b), so that the current is largely carried by the states within a transmission window $w = eV - (G_{FL}+G_{FR})/2$, where $G_{FL}$ and $G_{FR}$ are sizes of the Friedel gaps on the left and right sides, respectively. This window remains the same upon reversing bias polarity and no asymmetry arises. However, the bias voltage shifts both Friedel gaps downwards relative to the respective chemical potentials. One limiting case occurs when both Friedel gaps are completely located below their respective chemical potentials, as shown in the lower panel of Fig. 1b, where the transmission window becomes $w = eV - G_{FL}$. Upon bias polarity reversal, $w = eV - G_{FR}$. As a result, for the same current flowing under the reversed bias, there is a difference in the voltage drop across the interface by approximately $\Delta V = (G_{FL}-G_{FR})/e$.

Summarizing the above discussion, we see that when the bias is smaller than both gaps, there is no asymmetry (this limit ensures that there is no violation of the time reversal symmetry); when the bias is between the two gaps, the asymmetry increases linearly with the bias; when the bias is above both gaps, the asymmetry saturates to a constant $|G_{FL}-G_{FR}|/e$ (as explicitly shown in Section IV). Since Friedel gaps are opened only for normally incident electrons, a necessary condition for the transport asymmetry is that transmission is limited to the normal direction, which is satisfied in our experiments as described below.

## III. EXPERIMENT

### A. Experimental method

Our experiment employs monolayer-bilayer (ML-BL) interfaces in graphene formed on Si-face SiC (0001) [13-15]. The Si-face of SiC allows for better control of the graphene thickness than the C-face, due to the Si-face initially growing a buffer layer that acts as a template for the graphene formation as the Si is sublimed from the surface. To grow graphene, a 1 mm thick bow-tie shaped graphite heating plate with a narrow neck measuring about 14 mm × 20 mm was used to heat a 1 cm × 1 cm sample resting on the neck. This heater draws a current of ~200 A. Water-cooled copper clamps and electrical feedthroughs supply the current, and the heater is contained in an ultra-high-vacuum chamber. The SiC graphene growth procedure starts with hydrogen etching at 1620 °C for 3 min followed by the graphene growth at 1590 °C for 30 minutes in 1 atm argon environment.

We directly measure the voltage drop across a ML-BL graphene boundary by using a scanning tunneling potentiometry (STP) technique [16-18], implemented in a cryogenic multiple-probe scanning tunneling microscope (STM) [19, 20]. In the STP setup, schematically shown in Fig. 2a, two STM probes (probe 1 and probe 2) are in contact with the sample surface applying a constant current. A third tip (probe 3) is positioned between the current probes and scans the sample surface to measure both the topography and the local electrochemical potential ($\mu_{ec}$) at each point [18, 19, 21] at 80 K. Both sample and tip are maintained at same temperature.

Unlike conventional tunneling spectroscopy where the spectroscopic resolution is limited by thermal broadening of the electron energy distribution in the Fermi-distribution, the potentiometry technique measures the local electrochemical potential with nominally zero current flow at the tip-sample junction. In this case, the voltage noise in potentiometry is dominated by thermal noise that is $\Delta V = \sqrt{4 k_B T R_T \Delta f}$, where $\Delta f$ is the bandwidth [22]. A resolution better than 10 μV can be achieved at low temperatures [23].

## B. Measured transport asymmetry

Figures 2b and 2c show STM images of the ML and BL graphene, respectively, with a moiré pattern and atomic lattices. By comparing STM images and scanning tunneling spectroscopy (STS) acquired on both sides of the step [24], we find that the lattice structure of graphene remains unchanged across the boundary, indicating a carpet-like growth mode covering the substrate step and terrace, and the extra graphene layer underneath the graphene carpet primarily has an armchair type of edge structure. Epitaxial graphene on SiC (0001) is heavily n-doped due to charge transfer from a buffer layer [25], and a ML-BL boundary almost always coincides with a substrate step [26]. Therefore, a transition region of deformed graphene over the substrate step [27] is nearly undoped due to increased distance to the substrate [28]. This undoped region forms a barrier to incident electrons so that the transmission probability decreases sharply with transverse momentum, thus limiting transmission to near-normal incidence.

The STM image in Fig. 3a shows a ML-BL boundary, with a step height measured to be ~0.8 Å, in good agreement with the expectation from the interlayer spacing difference between SiC (2.5 Å) and BL graphene (3.3 Å) [29, 30]. Since the BL is slightly higher than the ML on the surface, we denote this transition as a "step-up" ML-BL boundary. Figure 3b is a schematic illustration of this boundary. Figure 3c shows potential profiles measured across this boundary for both bias polarities. A clear potential drop occurs at the step edge for each polarity. The potential drop at the boundary when the current flows from the BL to the ML ($V_-$, denoted as reverse bias) is clearly higher than when a current of precisely the same magnitude flows from ML to BL ($V_+$, denoted forward bias). On the same sample, we also identified "step-down" ML-BL boundaries, as shown in the STM image in Fig 3d and schematically illustrated in Fig. 3e. The STM measured step height of this "step-down" ML-BL boundary is 1.65Å. The same kind of bias reversal asymmetry, $V_- > V_+$, is observed, as shown in Fig. 3f.

For comparison, we measure the potential profiles across ML graphene covering a substrate step, referred to as a ML-ML "boundary" (Fig. 3g), corresponding to the situation depicted in Fig. 3h. The potential profiles measured at several different source current values are shown in Fig. 3i for both bias polarities. To facilitate comparison, the profiles measured at reverse bias are flipped and superimposed onto the corresponding forward bias profiles. Clearly, the potential drops are the same for forward and reverse biases. Thus, the ML-ML "boundary" exhibits symmetric transport, as expected for this homojunction. These results confirm that the transport asymmetry at the ML-BL boundary is intrinsic to the heterojunctions and exclude a substrate step-induced asymmetry scenario.

## IV. COMPARISON BETWEEN THEORY AND EXPERIMENT

To facilitate a quantitative comparison between theory and experiment, we explicitly estimate the size of $\Delta V$ for ML-BL graphene boundaries on SiC using the equations derived in the theory section and in Appendix A. The electron wave function has two components in ML graphene, and four in BL graphene. However, for the latter a two component form containing only the two dominant layer-sublattice pseudospins is often used [31, 32]. The Friedel oscillations in ML graphene are shown to be out of phase between the A and B sites [4, 33], i.e. $\varphi_A - \varphi_B = \pi$. We argue that this should be a more general result because the local Coulomb energy is minimized when the two oscillations are out of phase. Therefore the same consideration applies to the BL side where the phase between the two sublattices must also be out of phase. The amplitude of the Friedel oscillation decays away from the interface, thus the Friedel gap diminishes away from the interface, giving rise to a slanted effective potential barrier to normally incident electrons.

We now estimate the band gap in ML graphene. Considering $\rho_1^A = \rho_1^B = \rho_1$, the Friedel gap is simplified to $U\rho_1\Omega/2$ from Eq. (A3) in Appendix A. To estimate the parameter $\rho_1$ for the gap calculation, the $2k_F$ Fourier component of the charge density near a boundary is derived in Appendix A as being proportional to $x_M^{-3/2}$, where $x_M = \lambda_F/2$. Assuming that the charge oscillation amplitude is $\rho_0$ on each sublattice at the boundary, where $2\rho_0$ is the total charge density with $2\rho_0\Omega \approx 6\times10^{-3}$ electrons [34], we obtain $\rho_1 \approx \rho_0 (k_F x_M)^{-3/2} = \frac{\rho_0}{\pi^{3/2}} \approx 0.18\rho_0$. Since we have assumed the oscillation amplitude to be half the total charge density, the above estimate represents an upper bound. Using the onsite Coulomb energy $U \approx 9$ eV [35], the estimate for the gap on the ML side is $U\rho_1\Omega/2 = 2.43$ meV.

For BL graphene on SiC (0001), charge transfer from the buffer layer results in a vertical electric field [34], leading to pseudospin polarization [32], i.e. $|\alpha| \neq |\beta|$. According to San-Jose *et al.* [32], $\alpha = \sqrt{1-U_{12}/2(\mu-E_D)}$ and $\beta = -\sqrt{1+U_{12}/2(\mu-E_D)}$, where $U_{12}$ is the potential difference between the two atomic layers, $\mu - E_D$ the chemical potential measured from the Dirac point energy. Applying Eq. (A4) of the Appendix A for BL graphene, the Friedel gap is $U\rho_1\Omega U_{12} / 2(\mu - E_D)$. The amplitude of the charge density oscillation in the BL is estimated in the same manner as in the ML graphene, $\rho_1 \approx 0.18\rho_0$. Having $2\rho_0\Omega \approx 8.1 \times 10^{-3}$ electrons [34], $U_{12} \approx 0.11$ eV [30, 34], $\mu - E_D \approx 0.35$ eV (according to [30] and our own measurement [36]), and $U \approx 8$ eV [35], we obtain a Friedel gap of 0.92 meV for BL graphene. The difference between the ML and BL Friedel gaps is 1.5 meV. The theory thus predicts a polarity reversal asymmetry $\Delta V < \sim 1.5$ mV and a saturation voltage of ~2.4 mV.

The transport asymmetry, i.e. the difference in Friedel gaps in ML and BL graphene, originates from different chiralities of the two, as discussed in Appendix A. For out-of-phase charge density oscillations, the Friedel gap maximizes for the ML but minimizes for the BL. Had it not been for the pseudospin polarization in the BL, the Friedel gap would be zero on the BL side. Therefore, the observed asymmetry is a manifestation of the difference in chirality between ML and BL graphene.

Because the current probes only provide a total current, to avoid errors introduced in estimating the current density, we use the measured local voltage to quantify the transport asymmetry. We define $V_+ \equiv V(+|I|)$, $V_- \equiv |V(-|I|)|$, $\Delta V \equiv V_- - V_+$, and $\overline{V} \equiv (V_+ + V_-)/2$. Figure 4a shows $\Delta V$ vs. $\overline{V}$. At low biases ($\overline{V} < 1$ mV), there is no noticeable polarity reversal asymmetry. For higher bias, the observed asymmetry $\Delta V$ is mostly around 1 mV. Figure 4b shows the same data in a different view, plotting $V_-$ against $V_+$. Here, except those measured at biases < 0.5 mV, all data points obtained from five ML-BL boundaries fall on the same straight line with a slope of 1 and an intercept of about 1 mV. The inset to Fig. 4a shows the $\Delta V$ vs. $\overline{V}$ data points with $\overline{V}$

≤ 6 mV, along with theoretically estimates of the asymmetry depicted as the solid line. We note that Ji *et al* [21] carried out similar measurements at biases < 0.5 mV, and did not observe any asymmetry, consistent with both our theory and experimental data.

## V. SUMMARY AND DISCUSSION

In summary, we show that the Friedel oscillation at an interface opens an energy gap at the chemical potential. Although this gap only occurs near an interface for electrons with wave vectors perpendicular to the interface, it can play a key role in transport process with near-normal incidence. Under a bias voltage, the Friedel gaps on both sides of the interface shift downwards, and eventually sink completely below the respective chemical potentials when the bias is beyond a critical value. For a heterojunction, the Friedel gaps are different on the two sides, leading to asymmetric transport behavior upon bias polarity reversal. The polarity reversal asymmetry measured at ML-BL boundaries in epitaxial graphene on SiC (0001) is in strikingly good agreement with the theory, revealing the effect of Friedel gaps, which are difficult to measure directly since such a measurement must be angle-resolved and requires high energy resolution (sub-meV) as well as nanoscopic spatial resolution. Moreover, our theory and observation may provide a new avenue towards quantum manipulation of electron transport via chemical or electrostatic doping in graphene and topological insulators.

The transport asymmetry shown in Fig. 4 is in good agreement with our theoretical estimate, and stands in stark contrast to that of a typical nonlinear conductance induced by density of state mismatch or asymmetric transmission probability. As explained in detail in Appendix B, those nonlinear I-V curves can be expressed as a polynomial form $I = aV + bV^2 + O(V^3)$, which would lead to a reversal asymmetry of the form $\Delta V \propto \overline{V}^2$.

Without considering the Friedel gaps, electron transmission across ML-BL graphene boundaries is a smooth function of energy [31, 37-39]. Nonlinear transport in the polynomial form can be derived by considering, for example, the DOS mismatch between ML and BL graphene if there are no sharp bulk DOS features near the Fermi energy. But, our experimentally observed asymmetry is not of the form of $\Delta V \propto \bar{V}^2$.

In principle, a possible source of systematic error that cannot be excluded by our measurement of the ML-ML "boundary" is a thermovoltage change at the ML-BL junction due to the thermopower difference between the two sides [40]. To estimate the size of this error, we measured the thermovoltage across the ML-BL boundary in our cryogenic STP system (with a temperature gradient $\Delta T < 1$ K between the sample and STM tip) [41]. The measured thermovoltage value is less than 10 µV [41], 2 orders of magnitude smaller than the observed 1 mV transport asymmetry and hence unlikely to be the source of that asymmetry. Thermovoltage corresponds to the logarithmic derivative of the electronic density of states, and thus provides a good way to visualize local DOS variations [41]. Friedel oscillations associated with intra-valley scattering have indeed been observed on BL graphene in the thermovoltage distributions, but not on the ML graphene [41] as the oscillations associated with the two sublattices are opposite in phase and thus cancel each other, corroborating our above analysis.

Because the Friedel gap is proportional to both the onsite Coulomb energy parameter and the amplitude of the Friedel oscillation, the size of the effect can be enhanced by increasing either factor. Ohta et al [42] have demonstrated that the carrier density of graphene on SiC can be tuned in a wide range of $(5—35) \times 10^{-3}$ electrons per unit cell by chemical doping. Such a carrier density would lead to a gap size up to 14.2 meV on the ML side. The Friedel gap of the BL graphene will vanish if there is no pseudospin polarization, which is present in our sample due to charge transfer from the substrate. Therefore the difference between the two gaps, namely the magnitude of the transport asymmetry, can be up to 14.2 mV. The observed asymmetry is a manifestation of the Friedel oscillation on the ML side, which is usually unobservable. Furthermore, the ML and BL Friedel gaps are different primarily because graphene ML and BL possess different chiralities, therefore the observed asymmetry is also a manifestation of this chirality difference. Some topological insulators [43] exhibit very strong Coulomb interaction, indicating that they may be possible candidates for exploring Friedel gap effects. In contrast to graphene, the two sets of Friedel oscillations in topological insulators are associated with real spins instead of pseudospins on each sublattice, and a magnetic field at the boundary would be able to tune the phase of Friedel oscillations and thus switch on/off the Friedel gap.

**ACKNOWLEDGEMENT**S

This research was conducted at the Center for Nanophase Materials Sciences, which is sponsored at Oak Ridge National Laboratory by the Office of Basic Energy Sciences, U.S. Department of Energy. The work was partially supported by ORNL-UTK Joint Institute of Advanced Materials (JIAM) and the National Science Foundation.

**APPENDIX A: THEORETICAL CONSIDERATION FOR GRAPHENE**

**Friedel gaps in ML and BL graphene**

Friedel oscillations have been observed in both BL and ML graphene as quantum interference patterns in local density of states (DOS) [5, 6, 44-46]. Furthermore, as a result of the smooth scattering potential in the continuous top layer at ML-BL boundary on SiC (0001) [30, 44, 45, 47], only *intravalley* scattering induced long-wavelength Friedel oscillations can occur and *intervalley* scattering induced oscillations are absent [4, 41, 45]. Therefore the energy gap considered here is opened by the long-wavelength ($Q = 2k_F$) Friedel oscillation associated with intravalley scattering.

We need to consider a special case of a monolayer-bilayer (ML-BL) boundary where only intravalley scattering induced charge density oscillations are present. In this consideration, the K and K' valleys are equivalent, therefore only one needs to be considered. The two sublattices in graphene means that the wave function has two components, which are usually referred to as pseudo-spins, which in turn leads to two Friedel oscillation components with a common period but different phases. For monolayer (ML) graphene, the wave function can be written as [31]

$$|\mathbf{k}\rangle = \frac{1}{\sqrt{2\Omega}} \begin{pmatrix} \alpha \\ \beta \exp(i\theta) \end{pmatrix} \exp(i\mathbf{k} \cdot \mathbf{r}),$$

where $\boldsymbol{k} = \hat{\boldsymbol{x}}k_x + \hat{\boldsymbol{y}}k_y$ with $\theta = \arctan(k_y / k_x)$, $\Omega$ is the unit cell area (which contains both sublattice sites) over which the wave function is normalized, and $\alpha = \beta = 1$ in the absence of pseudo-spin polarization.

For BL graphene, we invoke the low-energy approximation for the wave function, which reduces the four-spinor (two sublattices on each sheet) into an effective two-spinor wave function [31, 32]:

$$|\mathbf{k}\rangle = \frac{1}{\sqrt{2\Omega}} \begin{pmatrix} \alpha \\ \beta \exp(i2\theta) \end{pmatrix} \exp(i\mathbf{k} \cdot \mathbf{r}),$$

where $\alpha$ and $\beta$ represent the two dominate pseudo-spins (sublattice A in one layer and sublattice B in the other) and $|\alpha| \neq |\beta|$ due to pseudo-spin polarization induced by the vertical field.

Since the transmission probability decreases sharply with transverse momentum $k_y$ thus limiting transmission to near-normal incidence due to the depletion region formed at the ML-BL boundary in our sample [36], we may consider only the normal incidence, i.e. $k_y = 0$. In order to have a unified expression for the Freidel gap that applies to both ML and BL graphene, we write for $k_x > 0$:

$$|+k_x\rangle = \frac{1}{\sqrt{2\Omega}}\begin{pmatrix}\alpha\\\beta\end{pmatrix}\exp(ik_x x)$$

and

$$|-k_x\rangle = \frac{1}{\sqrt{2\Omega}}\begin{pmatrix}\alpha\\\beta\exp(i\Theta)\end{pmatrix}\exp(-ik_x x),$$

for both ML and BL graphene, where $\alpha = \beta = 1$ and $\Theta = \pi$ for ML, while $|\alpha| \neq |\beta|$ and $\Theta = 2\pi$ for BL.

Next we derive a general two-component formula for the Friedel gap and then apply it to graphene. In the presence of an oscillating charge density on two sublattices,

$$\rho(x) = 2\rho_0 + \rho_1^A \cos(Qx + \varphi_A) + \rho_1^B \cos(Qx + \varphi_B),$$

the extra Coulomb term in the Hamiltonian is in the form,

$$\frac{1}{2}U\Omega\left[\rho_1^A\sum_{i_A}c_{i_A}^+ c_{i_A}\cos(Qx_{i_A} + \phi_A) + \rho_1^B\sum_{i_B}c_{i_B}^+ c_{i_B}\cos(Qx_{i_B} + \phi_B)\right], \tag{A1}$$

where $\rho_1^{A(B)}$ are the amplitudes of the oscillations on the two sublattices, $c_{i_{A(B)}}^+$ is the creation operator at lattice position $x_{i_{A(B)}}$ on sublattice $A(B)$, and $\varphi_{A(B)}$ are the phases of the oscillations. For BL graphene, A and B denote sublattice with A in one layer and sublattice B in the other, respectively. Without the extra Coulomb term, the pair of the right- and left-going wave functions that will be coupled eventually by the perturbation can be generally written as

$$\left|Q/2+q\right\rangle = \begin{pmatrix} \alpha \\ \beta \end{pmatrix} e^{i(Q/2+q)x} / \sqrt{2\Omega} \quad \text{and} \quad \left|-Q/2+q\right\rangle = \begin{pmatrix} \alpha \\ \beta\exp(i\Theta) \end{pmatrix} e^{-i(Q/2-q)x} / \sqrt{2\Omega} \quad \text{respectively.}$$

The part of the full Hamiltonian projected to these two states is,

$$\begin{pmatrix} E\left(\dfrac{Q}{2}\right) - cq & \dfrac{1}{8}U\Omega\left(\rho_1^A\alpha^2 e^{-i\phi_A} + \rho_1^B\beta^2 e^{-i(\phi_B-\Theta)}\right) \\ \dfrac{1}{8}U\Omega\left(\rho_1^A\alpha^2 e^{i\phi_A} + \rho_1^B\beta^2 e^{i(\phi_B-\Theta)}\right) & E\left(\dfrac{Q}{2}\right) + cq \end{pmatrix},$$

where we assume an (approximate) linear dispersion $E(Q/2\pm q) = E(Q/2)\pm cq$. This linear dispersion is rigorously true for ML graphene (where $c = \hbar v_F$, $v_F$ being the Fermi velocity of graphene). We find that new eigenstates are standing waves with energies

$$E = E_F \pm \frac{1}{8}U\Omega\sqrt{\left(\rho_1^A\alpha^2\right)^2 + \left(\rho_1^B\beta^2\right)^2 + 2\rho_1^A\rho_1^B\alpha^2\beta^2\cos(\phi_A-\phi_B+\Theta) + (8cq)^2} \ . \qquad \text{(A2)}$$

The angle $\Theta$ is commonly referred to as the Berry's phase. For ML graphene, $\Theta = \pi$, thus the gap is maximum when the oscillations on two sublattices are out of phase to each other. For BL graphene, $\Theta = 2\pi$, so the gap is maximum when the oscillations on two sublattices (in this case, in two separate sheets) have the same phase. In general the Coulomb energy is minimized when the oscillations on two sublattices are out of phase. Indeed, there seems to be a consensus that in ML graphene this is the case and consequently Friedel oscillation cannot be observed in ML graphene [4, 33]. The same energy minimization requirement should also lead to a similar result for BL graphene but because of the difference between $\alpha$ and $\beta$, the cancellation is not complete so that the oscillation should be observable in BL. While the charge oscillations are always minimized by the Coulomb energy, the different Berry's phases in the ML and BL systems lead to opposite effect on the Friedel gap. For the same out-of-phase oscillations on two sublattices, the gap in the ML is maximized while the gap in the BL is minimized. Therefore the transport asymmetry at the graphene boundary due to the difference in the Friedel gap on both sides is fundamentally linked to the different chirality of the ML and BL systems.

Imposing the condition $\varphi_A - \varphi_B = \pi$, and for ML setting $\alpha = \beta = 1$ and $\rho_1^A = \rho_1^B = \rho_1$, we obtain,

$$E = E_F \pm \frac{1}{4} U\Omega \sqrt{\rho_1^2 + (4cq)^2} \ . \tag{A3}$$

For BL graphene, we note that $\rho_1^A + \rho_1^B = 2\rho_1$ but $\rho_1^A / \rho_1^B = \alpha^2 / \beta^2$ and $\alpha^2 + \beta^2 = 2$. Thus $\rho_1^A = \alpha^2 \rho_1$ and $\rho_1^B = \beta^2 \rho_1$. The band dispersion is found as,

$$E = E_F \pm \frac{1}{4} U\Omega \sqrt{\left(\alpha^2 - \beta^2\right)^2 \rho_1^2 + (4cq)^2} \ . \tag{A4}$$

**Amplitude decay of Friedel oscillations in ML and BL graphene**

In the above discussion, as well as in Section I, we treated the Friedel oscillation as a periodic potential without considering its decay. Now we consider the decay rate for the Friedel oscillation amplitude. The change in the local density of states due to the scattering off the boundary is

$$\delta\rho_{A(B)}\left(E, x\right) = \frac{k}{\hbar v_F} \int_{-\pi/2}^{\pi/2} d\theta \frac{4|r|}{1 + |r|^2} \cos\left(2kx\cos\theta - \Phi_{A(B)}\right), \tag{A5}$$

following the method in Ref. [48], and we include the graphene density of states in the prefactor which was omitted by Ref. [48]. Here, $\Phi_{A(B)}$ is the phase of the reflection coefficient $r$ for each sub-lattice and should in principle depend on $\theta$. In our case the boundary is strongly reflective [36] so $|r| \approx 1$ and we cannot use $r \approx \sin\theta$ as in Ref. [48]. To capture the leading even and odd terms of $\Phi_{A(B)}(\theta)$, we note that the dominant contribution to the Friedel oscillation is from small $\theta$ and write

$$\Phi_{A(B)} \approx -\varphi_{A(B)} - \frac{3}{4}\pi + C_{A(B)}\theta. \qquad (A6)$$

The negative sign and the extra phase of $-\frac{3}{4}\pi$ are included to ensure a consistent phase definition with Eq. (A1), as we will see below. The derivation below shows that inclusion of higher order terms in $\theta$ does not change the result other than an overall scaling constant. Inserting (A6) into (A5), we have

$$\delta\rho_{A(B)}(E,x) = \frac{k}{\hbar v_F} \int_{-\pi/2}^{\pi/2} d\theta \cos\left(C_{A(B)}\theta\right)\cos\left(2kx\cos\theta + \phi_{A(B)} + \frac{3}{4}\pi\right).$$

The total Friedel charge density is obtained by integrating the local density of states over the energy,

$$\delta\rho_{A(B)}(x) = \int_{E_D}^{E_F} dE \frac{k}{\hbar v_F} \int_{-\pi/2}^{\pi/2} d\theta \cos\left(C_{A(B)}\theta\right)\cos\left(2kx\cos\theta + \phi_{A(B)} + \frac{3}{4}\pi\right).$$

For ML, $E = E_D + \hbar v_F k$,

$$\delta\rho_{A(B)}(x) = \int_0^{k_F} kdk \int_{-\pi/2}^{\pi/2} d\theta \cos\left(C_{A(B)}\theta\right)\cos\left(2kx\cos\theta + \phi_{A(B)} + \frac{3}{4}\pi\right). \qquad (A7)$$

For BL, the energy dispersion is not linear. However, it is approximately linear for energies sufficiently away from the band bottom and the contribution to the Friedel oscillation from energies close to the band bottom is much smaller. Therefore, the linear dispersion can be approximately used for the entire integration range even for the BL. Integrating over $k$ first, we find,

$$\delta\rho_{A(B)}(x) = \int_{-\pi/2}^{\pi/2} d\theta \cos\left(C_{A(B)}\theta\right)\left[\frac{\sin\left(2k_F x\cos\theta + \phi_{A(B)} + \frac{3}{4}\pi\right)}{2x\cos\theta}k_F\right.$$

$$\left. + \frac{\cos\left(2k_F x\cos\theta + \phi_{A(B)} + \frac{3}{4}\pi\right) - \cos\left(\phi_{A(B)} + \frac{3}{4}\pi\right)}{\left(2x\cos\theta\right)^2}\right]$$

We apply the stationary phase approximation to integrate over $\theta$, and get the leading oscillatory term as,

$$\delta\rho_{A(B)}(x) \propto \frac{\cos\left(2k_F x + \varphi_{A(B)}\right)}{\left(k_F x\right)^{3/2} x}.$$

This is the expected $x^{-5/2}$ decay far away from the boundary, but the Friedel gap is determined by the $2k_F$ Fourier components of the oscillation at small distances. To find these components, we apply the Fourier transform between $x = 0$ and $x = x_M$ (which we will define as half of the Fermi wavelength since any barrier less than half a Fermi wavelength in width cannot effectively block the transmission) to Eq. (A7) and obtain,

$$\frac{1}{x_M}\int_0^{x_M}\delta\rho_{A(B)}(x)e^{i2k_F x}dx = \frac{1}{x_M}\int_0^{k_F}kdk\int_{-\pi/2}^{\pi/2}d\theta\cos\left(C_{A(B)}\theta\right)\int_0^{x_M}\cos\left(2kx\cos\theta + \phi_{A(B)} + \frac{3}{4}\pi\right)e^{i2k_F x}dx,\text{ with}$$

$$\int_0^{x_M}\cos\left(2kx\cos\theta + \phi_{A(B)} + \frac{3}{4}\pi\right)e^{i2k_F x}dx = \frac{1-e^{i2(k_F+k\cos\theta)x_M}}{4(k_F+k\cos\theta)}ie^{i\left(\phi_{A(B)}+\frac{3}{4}\pi\right)} + \frac{1-e^{i2(k_F-k\cos\theta)x_M}}{4(k_F-k\cos\theta)}ie^{-i\left(\phi_{A(B)}+\frac{3}{4}\pi\right)}.$$

The contribution to the Fermi surface integral is dominated by the region around $k \approx k_F$ and $\cos\theta \approx 1$, where the leading term of the above expression is approximately,

$$\frac{1}{2}x_M e^{-i\left(\phi_{A(B)}+\frac{3}{4}\pi\right)}e^{i2(k_F-k\cos\theta)x_M}.$$

This yields the leading term of the Fourier component as,

$$\frac{1}{2}e^{-i\left(\phi_{A(B)}+\frac{3}{4}\pi\right)}\int\limits_{0}^{k_F}kdk\int\limits_{-\pi/2}^{\pi/2}d\theta\cos\left(C_{A(B)}\theta\right)e^{i2(k_F-k\cos\theta)x_M}=$$

$$\frac{1}{2}e^{-i\left(\phi_{A(B)}+\frac{3}{4}\pi\right)}\int\limits_{-\pi/2}^{\pi/2}d\theta\cos\left(C_{A(B)}\theta\right)\left[\frac{e^{i2(1-\cos\theta)k_F x_M}}{2ix_M\cos\theta}k_F+\frac{e^{i2(1-\cos\theta)k_F x_M}-1}{\left(2ix_M\cos\theta\right)^2}\right]$$ .

Applying the stationary phase approximation to this integral, we find the leading term of the Fourier component to be proportional to $x_M^{-3/2}$ .

## APPENDIX B: EFFECTS OF DENSITY OF STATE MISMATCH

We consider here the effects of a density of state (DOS) mismatch in the two electrodes of a tunnel junction, as well as possible asymmetric transmission probabilities under opposite biases. In general, a nonlinear current-voltage (*I-V*) relation of the tunnel junction can be expressed as

$$I = aV + bV^2 + O(V^3) .\tag{B1}$$

The second order term leads to asymmetry under polarity reversal. To quantify the asymmetry, we define $\Delta I \equiv |I(V)| - |I(-V)|$ and $\bar{I} \equiv [|I(V)| + |I(-V)|]/2$ when measuring $I(V)$ and $I(-V)$, where small $V > 0$. Similarly, $\Delta V \equiv |V(I)| - |V(-I)|$ and $\bar{V} \equiv [|V(I)| + |V(-I)|]/2$, if the voltages are measured at current biases $I$ and $-I$. For small $\bar{V}$, the polarity reversal asymmetry due to a nonlinear *I-V* relation described by Eq. (B1) can be quantified by

$$\frac{\Delta I}{\bar{I}} = -\frac{\Delta V}{\bar{V}} = \frac{2b}{a}\bar{V} .\tag{B2}$$

We consider tunneling currents that can be written as

$$I \propto \int_{-\infty}^{\infty} T(E)D_2(E)D_1(E)[f_2(E) - f_1(E)]dE ,\tag{B3}$$

where $T(E)$ is the transmission probability, and $D_2(E)$ and $D_1(E)$ are DOS functions of the two sides of the junction. For simplicity, we consider the low temperature limit

$$I \propto \int_{\mu}^{\mu+eV} D_1(E) T_V(E) D_{20}(E - eV) dE .$$

(B4)

Here, $T_V(E)$ is the transmission probability of energy level $E$ at a bias $V$, and $D_{20}(E)$ is the DOS on side 2 at zero bias, and $\mu$ is the chemical potential on side 1. Any possible asymmetry of the transmission probability under $\pm V$ is captured by the $V$ dependence of $T_V(E)$. In the following, we will drop the 0 in the subscript of $D_{20}$ for ease of notation.

To evaluate Eq. (B4) to the second order of $V$, we write $T_V(E)$ to the first order as $T_V(E) = T_0(E) + KeV$, where $T_0(E)$ is the transmission probability at zero bias, and $K$ is a proportional constant. We can further expand $T_0(E)$ in the neighborhood of $\mu$ and have

$$T_V(E) = T_0(\mu) + T_0'(\mu)(E - \mu) + KeV ,$$

(B5)

where $T_0'(\mu) \equiv \left. \dfrac{dT_0(\varepsilon)}{d\varepsilon} \right|_{\varepsilon=\mu}$ . We also expand $D_1(E)$ and $D_2(E - eV)$ around $\mu$, and have

$$D_1(E) = D_1(\mu) + D_1'(\mu)(E - \mu),$$

(B6)

$$D_2(E - eV) = [D_2(\mu) - D_2'(\mu)eV] + D_2'(\mu)(E - \mu) .$$

(B7)

Inserting Eqs. (B5) through (B7) into (B4), we get the integrand of Eq. (B4) to the first order of $(E - \mu)$ as:

$$D_1(\mu)[T_0(\mu) + KeV][D_2(\mu) - D_2'(\mu)eV] + \{D_1(\mu)[T_0(\mu) + KeV]D_2'(\mu) +$$
$$[T_0'(\mu)D_1(\mu) + (T_0(\mu) + KeV)D_1'(\mu)][D_2(\mu) - D_2'(\mu)eV]\}(E - \mu)$$

Take the integral, and we have

$$I \propto D_1(\mu)T_0(\mu)D_2(\mu)eV + [D_1(\mu)D_2(\mu)K + \frac{1}{2}D_1(\mu)T_0'(\mu)D_2(\mu)$$
$$+ \frac{1}{2}D_1'(\mu)T_0(\mu)D_2(\mu) - \frac{1}{2}D_1(\mu)T_0(\mu)D_2'(\mu)](eV)^2 + O(e^3V^3)$$

(B8)

A good approximation for $T_V(E)$ can be

$$T_V(E) = T_0(E - eV/2) = T_0(E) - T_0'(E)eV/2 . \tag{B9}$$

Comparing Eq. (B9) to $T_V(E) = T_0(E) + KeV$, we have $K = -T_0'(\mu)$ in the neighborhood of $\mu$. Inserting into Eq. (B8), we get

$$I \propto D_1(\mu)T_0(\mu)D_2(\mu)eV + \frac{1}{2}T_0(\mu)[D_1'(\mu)D_2(\mu) - D_1(\mu)D_2'(\mu)](eV)^2 + O(e^3V^3) \tag{B10}$$

If the junction is symmetric, the second order term vanishes.

To examine the bias polarity reversal asymmetry associated with this nonlinear *I-V* relation, we comparing Eq. (B10) and (B1) to yield

$$a = eD_1(\mu)T_0(\mu)D_2(\mu)e, \text{ and } b = \frac{e^2}{2}T_0(\mu)[D_1'(\mu)D_2(\mu) - D_1(\mu)D_2'(\mu)] .$$

Using Eq. (B2), we get

$$\frac{\Delta V}{\overline{V}} = -\frac{2b}{a}\overline{V} = -e\frac{D_1'(\mu)D_2(\mu) - D_1(\mu)D_2'(\mu)}{D_1(\mu)D_2(\mu)}\overline{V} . \tag{B11}$$

The above analysis shows that the nonlinear *I-V* relation of a tunneling junction is described by Eq. (B1), and the associated polarity reversal asymmetry can be calculated using Eq. (B11). An interesting case, where one of the two DOS functions, say, $D_1$ is constant, leads to a very simple expression

$$\frac{\Delta V}{\overline{V}} = e\frac{D_2'(\mu)}{D_2(\mu)}\overline{V} . \tag{B12}$$

We apply Eq. (B12) to the ML-BL graphene boundary on SiC(1000). The coincident step in the SiC substrate leads to a barrier. We therefore take the approximation that only normally incident electrons can tunnel through the barrier. In 1D (normal incidence), the linear $E(k)$ dispersion of ML graphene leads to a constant DOS $D_1$, therefore Eq. (B12) applies.

To find the BL graphene DOS in 1D, $D_2(E)$, we first write down the BL graphene band dispersion [49]:

$$E = \sqrt{(\hbar v_F)^2 k^2 + (\gamma_1/2)^2} - \gamma_1/2, \tag{B13}$$

where $v_F \approx 10^8$ cm/s, and $\gamma_1 \approx 0.4$ eV. In 1D, we have

$$D_2(E) \propto \frac{dk}{dE} = \frac{\sqrt{(\hbar v_F)^2 k^2 + (\gamma_1/2)^2}}{(\hbar v_F)^2 k}. \tag{B14}$$

Using Eq. (B14) and after some lengthy algebra, we get

$$\frac{D_2'(\mu)}{D_2(\mu)} = \frac{1}{\mu + \dfrac{\gamma_1}{2}} - \frac{\mu + \dfrac{\gamma_1}{2}}{\left(\mu + \dfrac{\gamma_1}{2}\right)^2 - \left(\dfrac{\gamma_1}{2}\right)^2}. \tag{B15}$$

Due to charge transfer from the substrate, the chemical potential referenced to the charge neutral energy for BL graphene on SiC(0001) is about $\mu = 0.35$ eV, which yields $\dfrac{D_2'(\mu)}{D_2(\mu)} = -0.28$ eV$^{-1}$. By Eq. (B12), we have

$$\Delta V = -(0.28\,\text{V}^{-1})\bar{V}^2. \tag{B16}$$

The sign indicates that a smaller potential drop will be measured when electrons tunnel from BL to ML (i.e. the current flows from the ML to BL). For typical STP measurements, the potential jump at the junction, $\bar{V}$, is on the order of mV, therefore the asymmetry is minuscule. In the STP performed by Ji *et al* [21], $\bar{V} < 0.3$ mV, therefore $-\Delta V < 0.03$ µV. The estimate is consistent with the absence of observable asymmetry in their measurements. When $\bar{V} = 20$ mV, however, we have $|\Delta V| = 0.1$ mV, which should start to become observable in careful STP measurements.

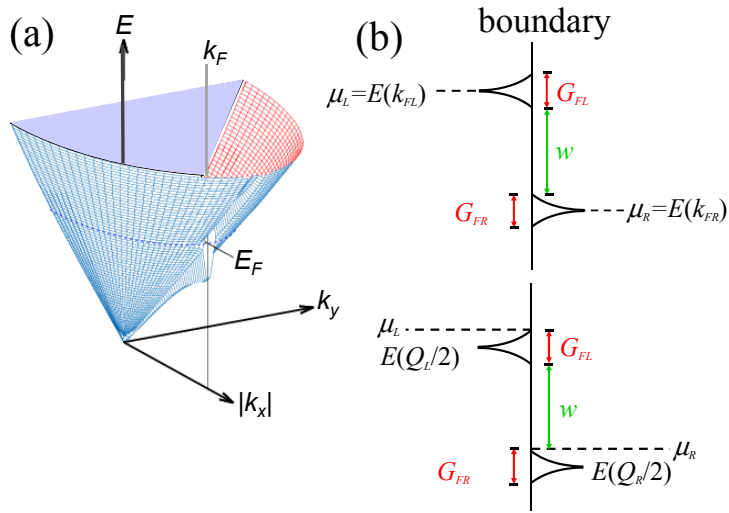

Fig. 1 (color online). (a) Schematic illustration of the Friedel gap at $E_F$ for wave vectors perpendicular to an interface, assuming a linear unperturbed dispersion (e.g. that of graphene). (b) Schematics showing the Friedel gap shifting on two sides of an interface. Upper panel: Friedel gaps open at the respective chemical potentials, without considering the nonequilibrium effect of Friedel gap shifting. Lower panel: both gaps completely exist below their respective chemical potentials.

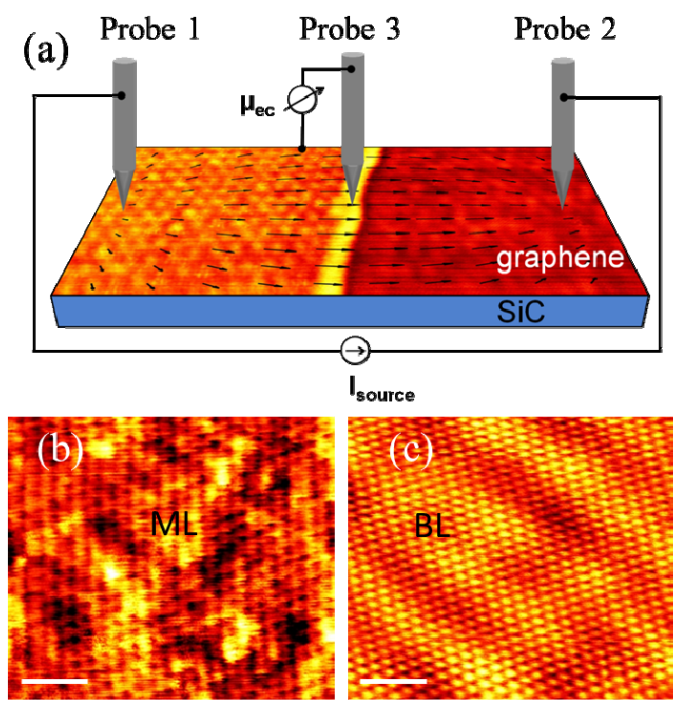

Fig. 2 (color online). (a) Schematic of the STP measurement setup. (b) STM images of ML and

(c), BL graphene. Scale bar: 1 nm.

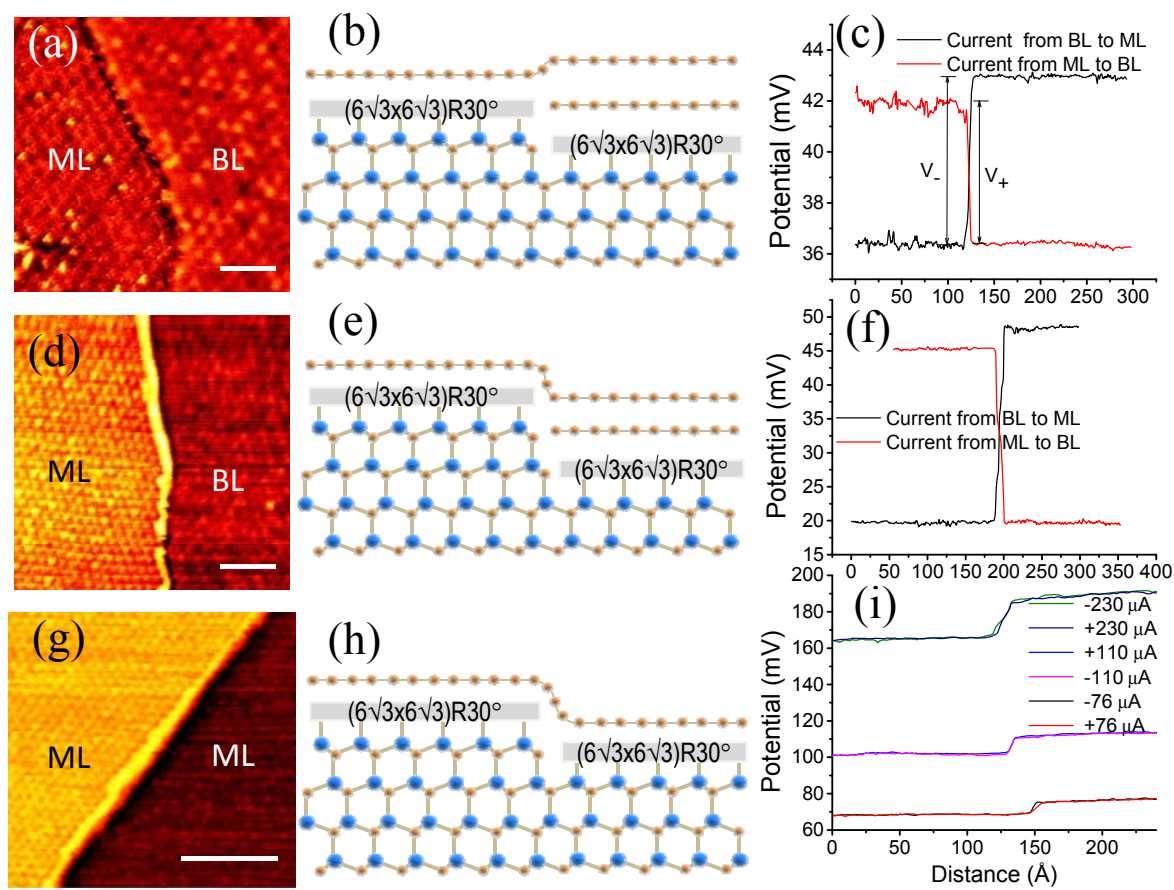

Fig. 3 (color online, 2 columns). (a) STM image of a step-up boundary. Scale bar: 6 nm. (b) Schematic of the step-up boundary. (c) Potential profiles measured across the boundary with forward and reversed bias conditions (222 μA current). (d) STM image of a step-down boundary. Scale bar: 10 nm. (e) Schematic of the step down boundary. (f) Potential profiles measured across the boundary with forward and reversed bias conditions (313 μA current). (g) STM image of a ML-ML graphene "boundary". Scale bar: 15 nm. (h) Schematic of a ML-ML boundary. (i) Potential profiles measured across the boundary with different current values and directions.

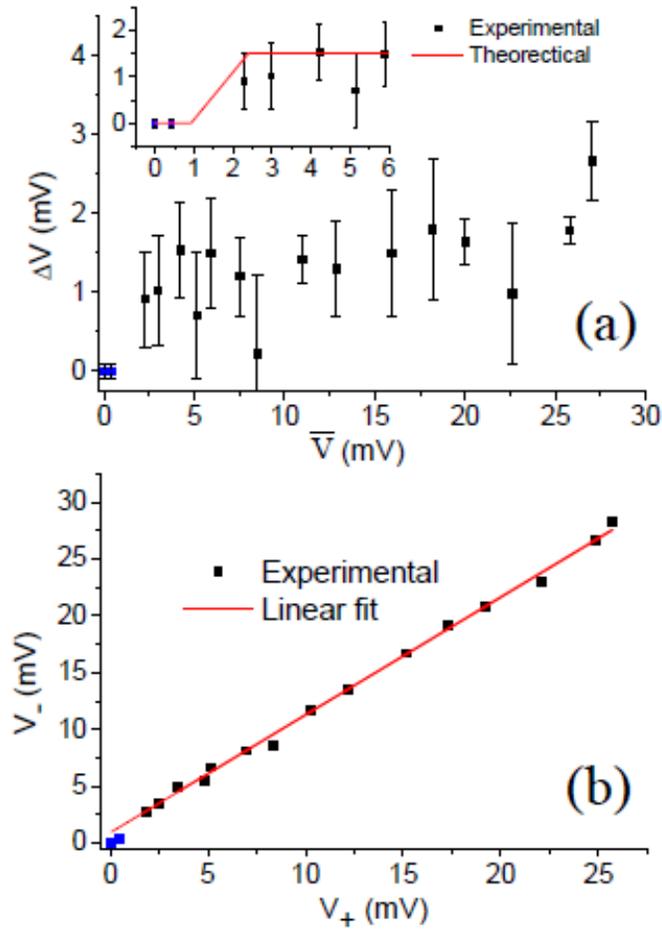

Fig. 4 (color online). (a) The asymmetry of potential drops ($\Delta V$) as a function of the averaged potential drop ($\overline{V}$) at the junction between forward and reverse biases. Inset: measured and calculated upper bound of $\Delta V$ at low $\overline{V}$. (b) The potential drop at reverse bias ($V_-$) as a function of potential drop at forward bias ($V_+$) at the junction. Data points shown in blue color correspond to those measured with a bias voltage less than 0.5 mV.